\documentclass[a4paper]{spie}  
\usepackage[]{graphicx}
\usepackage{amssymb,amsmath,psfrag,url}
\usepackage{verbatim}

\title{Method for fast computation of angular light scattering spectra from 2D periodic arrays}

\author{
Jan~Pomplun,\supit{\,a}
Sven~Burger,\supit{\,ab}
Lin~Zschiedrich,\supit{\,a}
Philipp Gutsche,\supit{\,ab}
Frank~Schmidt\supit{\,ab}
\skiplinehalf
\supit{a}
JCMwave GmbH,
Bolivarallee~22, 
D\,--\,14\,050 Berlin,
Germany
\smallskip\\
\supit{b}
Zuse Institute Berlin\,(ZIB),
Takustra{\ss}e~7,
D\,--\,14\,195 Berlin,
Germany
\authorinfo{
Corresponding author: S.~Burger\\
URL: http://www.jcmwave.com\\
URL: http://www.zib.de
}}

\begin{document}
\maketitle
\noindent
This paper will be published in Proc.~SPIE Vol.~{\bf 9778}
(2016) 977839 ({\it Metrology, Inspection, and Process Control for Microlithography XXX}, DOI: 10.1117/12.2219666)
and is made available 
as an electronic preprint with permission of SPIE. 
One print or electronic copy may be made for personal use only. 
Systematic or multiple reproduction, distribution to multiple 
locations via electronic or other means, duplication of any 
material in this paper for a fee or for commercial purposes, 
or modification of the content of the paper are prohibited.
Please see original paper for images at higher resolution. 

\begin{abstract}
An efficient numerical method
for computing angle-resolved light scattering off periodic arrays is 
presented. 
The method combines finite-element discretization with a Schur complement solver. 
A significant speed-up of the computations in comparison to standard finite-element 
method computations is observed. 
\end{abstract}



\section{Introduction}
Optical metrology is used in semiconductor and photomask manufacturing processes to determine 
a variety of characteristics of patterned wafers or photomasks such as layer thicknesses, 
width and shape of patterned structures, material properties and distribution, 
or overlay between different layers or structures.
State-of-the-art methods determining these features include spectroscopic 
or angle-resolved scatterometry, ellipsometry, reflectometry or imaging based methods~\cite{Pang2012aot}. 
These methods consist of an optical system which illuminates the measured sample 
and a detector system which measures properties of the light after interaction 
with the sample, e.g., the intensity, phase and/or polarization state of reflected and/or transmitted, resp.~of scattered light. 
For interpretation of the measurement, the measurement data are related to 
results obtained from modeling and simulation 
of the measurement setup for different values of the sample characteristics. 
These modeled distributions can be computed during the measurement process 
or can be precomputed in advance and stored in libraries. 
For a large number of free parameters, computation of these modeled light distributions can be 
very demanding.
Free parameters are physical characteristics of the modeled sample, such as geometrical or 
material properties. Further, a number of parameters is needed to describe the illumination system, 
e.g., multiple wavelengths and incidence directions or shape and polarization of the light beam. 
The incident light field generated by the illumination system is typically a beam with 
a finite numerical aperture, incident from a specific direction, or it has a complex shape, 
e.g., in quadrupole or freeform illumination. 
Such light distributions are typically modeled by a superposition of a number of fields, 
e.g., plane waves, with different incident directions, phases, and intensities~\cite{Wong2005a}.
The number of these fields needed for accurate modeling can be rather large (e.g., hundreds of fields).

If the measured sample is modeled as a periodic pattern, rigorous computation methods
like the Finite Element Method~\cite{MON03} (FEM), Rigorously Coupled Wave Analysis (RCWA) 
or Finite Difference Time Domain Method (FDTD) generally have to compute a solution separately 
for each incident direction. 
In addition to the variable physical characteristics, this significantly increases time and effort 
to generate the necessary modeled light distributions. 
The case of periodic patterns is highly important in semiconductor manufacturing and other technologies since 
many fabricated structures exhibit periodicity. 
Similar requirements for computations of solutions to Maxwell's equations
as in the above described field of computational metrology are present, e.g., in the 
field of computational lithography~\cite{Lai2012aot}.

A need therefore exists for a method to speed-up computation 
of such light distributions incident from a number of different directions on periodic samples.
In this contribution we present results from an implementation of a Schur complement based method~\cite{Golub2012matrix} 
allowing for such speed-ups. 
We demonstrate results that are based on JCMsuite, a FEM solver for
the linear Maxwell's equations in frequency domain. 
The implementation contains features like higher-order edge-elements~\cite{Pomplun2007pssb},
domain-decomposition~\cite{Schaedle_jcp_2007}, hp-adaptivity~\cite{Burger2015al}, and
model order reduction\cite{Pomplun2010SIAM}.
It is applied to tasks in computational lithography and metrology 
\cite{Scholze2008a,Pomplun2010bacus,Kleemann2011eom3,Petrik2015jeos} 
as well as to further fields in technology and 
research, see, e.g., for an overview~\cite{Burger2013pw}.

This paper is structured as follows: 
The background of the numerical method is presented in Section~\ref{section_background}. 
Section~\ref{section_example} presents simulation results on computation of angle resolved reflection spectra from 
2D-periodic arrays of silicon nano-posts.


\section{Schur complement method}
\label{section_background}
In this contribution we present a method to efficiently perform computation of scattered light fields  
from a periodic sample for a number of incident directions. 
The computational effort scales only very slowly with the number of incident directions. 
This can lead to substantial savings up to orders of magnitude in time and hardware costs 
for the computer system of a metrology or computational lithography system.

The general idea of the presented method is the decomposition of the light scattering model 
into a part which is independent on the direction of the incident field and a part which 
depends on the direction of the incident field. 
The part which is independent on the incident direction is solved once for all directions. 
The part dependent on the incident direction is solved for each incident direction individually. 
The individual and the common solutions are combined to give the total solution.

As a prerequisite let us look at a linear system of equations (as resulting, e.g., from FEM discretization 
of Maxwell's equations~\cite{Pomplun2007pssb}):
\begin{eqnarray}
  Ax&=&f,\label{eqBase}
\end{eqnarray}
where $A$ is a matrix, $f$ is the right hand side,  and $x$ is the sought solution.

Now we split the matrix into following blocks:
\begin{eqnarray}\label{eqBlocks}
  A&=&\left(
\begin{array}{cc}
A_{11} & A_{12}\\
A_{21} & A_{22}
\end{array}
\right),
\end{eqnarray}
where $A_{11}$ and $A_{22}$ are square matrices of generally different dimensions. 
Further, we split the solution and right hand side into corresponding blocks:
\begin{eqnarray}
  x&=&\left(
\begin{array}{c}
x_1\\
x_2
\end{array}
\right),\\
  f&=&\left(
\begin{array}{c}
f_1\\
f_2
\end{array}
\right).
\end{eqnarray}
Using this decomposition, Eq. \eqref{eqBase} can be written as follows:
\begin{eqnarray}
A_{11}x_1+A_{12}x_2&=&f_1\label{eqSplit1}\\
A_{21}x_1+A_{22}x_2&=&f_2.\label{eqSplit}
\end{eqnarray}
Equation \eqref{eqSplit1} can be solved with respect to $x_1$:
\begin{eqnarray}
x_1&=&A_{11}^{-1}\left(f_1-A_{12}x_2\right)\label{eqSchur1}
\end{eqnarray}
and used in \eqref{eqSplit}, giving:
\begin{eqnarray}
\left(A_{22}-A_{21}A_{11}^{-1}A_{12}\right)x_2=f_2-A_{21}A_{11}^{-1}f_1. \label{eqSchur}
\end{eqnarray}
Matrix $\left(A_{22}-A_{21}A_{11}^{-1}A_{12}\right)$ is the so-called Schur complement~\cite{Golub2012matrix} of $A$.

Now let us consider typical light scattering models, arising from discretization of 
Maxwell's equations, using e.g., FEM, FDTD, 
Discontinuous Galerkin, Finite Volume Method, or Volume Integral Method. 
The scattering model from a periodic structure is thereby formulated on a single 
unit cell and periodic boundary conditions are applied. 
These periodic boundary conditions depend on the grid vectors of the periodic domain and 
the incidence direction of the incoming light field, i.e., the phase difference from opposite 
periodic boundaries. 
This means that a number of entries in matrix $A$ depend on the incident direction and hence have to be solved independently.

\begin{figure}[b]
\begin{center}
\psfrag{cd1}{\small \sffamily CD$_1$}
\psfrag{cd2}{\small \sffamily CD$_2$}
\psfrag{cd3}{\small \sffamily CD$_3$}
\psfrag{h1}{\sffamily $h_1$}
\psfrag{h2}{\sffamily $h_2$}
\psfrag{px}{\sffamily $p_x$}
\psfrag{py}{\sffamily $p_y$}
\psfrag{air}{\sffamily air}
\psfrag{silicon}{\sffamily silicon}
  \includegraphics[width=.32\textwidth]{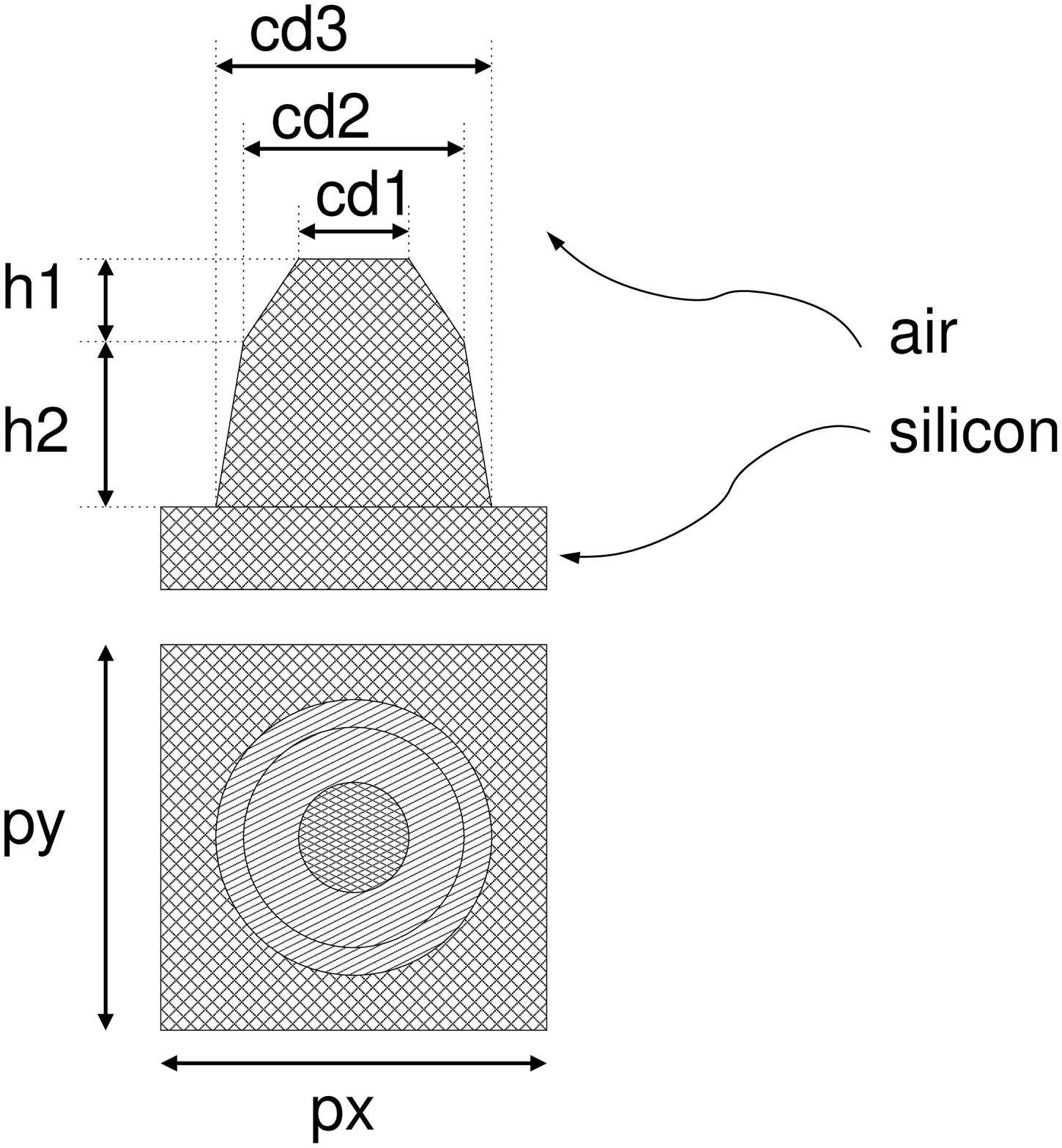}
\hspace{5mm}
  \includegraphics[width=.36\textwidth]{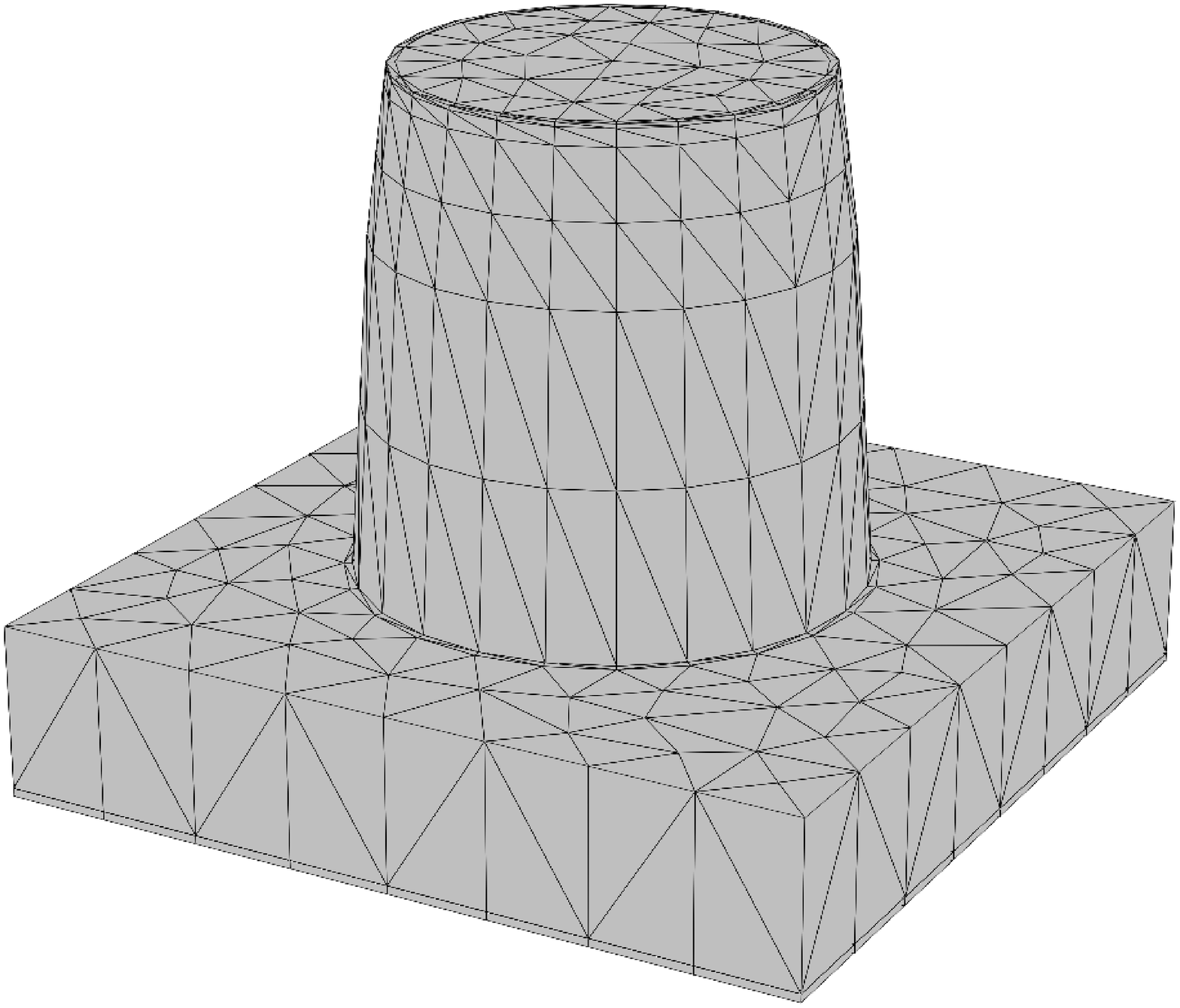}
  \caption{
{\it Left:} Schematics of the geometry of the investigated scattering geometry (unit cell of a 2D-periodic grating of nano-rods with 
circular cross section. 
Parameters of the model are indicated (critical dimension, CD, at different heights, $h_1$, $h_2$, pitches $p_x$, $p_y$), 
cf., Table~\ref{table_specs_nanopost}. 
{\it Right:} Visualization of the mesh discretizing the geometry (only silicon parts shown). 
}
\label{schematics_nanopost}
\label{fig_nanopost_rsp}
\end{center}
\end{figure}

In order to use above Schur splitting we reorder the linear system, 
such that all entries independent on the incident direction $\theta$ are in block $A_{11}$:
\begin{eqnarray}
A_{11}x_1+A_{12}(\theta)x_2&=&f_1(\theta)\\
A_{21}(\theta)x_1+A_{22}(\theta)x_2&=&f_2(\theta).\label{eqSplit2}
\end{eqnarray}
Hence, looking at \eqref{eqSchur1} and \eqref{eqSchur} we observe that inversion 
of block $A_{11}$ has to be performed only once, independent on the incidence direction. 
Only the solution of \eqref{eqSchur} is needed for each incidence direction. 

Typically the size of $A_{11}$ is much larger than the size of $A_{22}$ since 
the size of block $A_{22}$ only corresponds to the degrees of freedom on the 
boundary of the periodic computational domain. 
These are much less than the number of degrees of freedom in the volume in block $A_{11}$.

In order to arrive at the Schur system \eqref{eqSchur} the inverted 
(i.e., LU-decomposed) matrix is used for solving following systems:
\begin{eqnarray}
A_{11}^{-1}A_{12}\left(\theta\right)&\\
A_{11}^{-1}f_1\left(\theta\right),
\end{eqnarray}
and the solutions are used in \eqref{eqSchur} to assemble the $\theta$ dependent Schur complement:
\begin{eqnarray}
A_{22}\left(\theta\right)-A_{21}\left(\theta\right)A_{11}^{-1}A_{12}\left(\theta\right)
\end{eqnarray}

 Once this dense Schur system is solved for $x_2$, the inner solution $x_1$ can be obtained from \eqref{eqSchur1}. 
Also here, the decomposition of $A_{11}$ can be used for all incidence directions. 
An increasing number of incidence directions only leads to an increasing number of right hand sides. 
Especially for discretization of 3D scattering problems, 
however the LU-decomposition of the system takes much more effort than solving 
for a large number of right hand sides.

In the following example we will show that the Schur splitting can lead to substantial 
savings in computation time when scattering problems with a large number of right hand 
sides (i.e., sources with various incidence directions) are solved. 


\section{Application example: \newline Light scattering off an array of nano-posts}
\label{section_example}
In this section, the computation of angle-resolved scattering spectra is analyzed 
as an example of the described Schur complement technique.
Corresponding setups are relevant for 
applications like, e.g., Fourier ellipsometry~\cite{Petrik2015jeos}.
The following model is investigated:
Light with a vacuum wavelength of $\lambda_0 = 300\,$nm and well-defined polarizations (S- and P-polarized) 
illuminates a 3D structure. The source is modeled as a series of independent, 
coherent plane waves with well defined wave-vectors $\vec{k}=(k_x, k_y, k_z)$.
For normal incidence, the sample is arranged such that $\vec{k}=(0,0, k_z)$, $k_z=2\pi/\lambda_0$.
The geometry of the illuminated structure is a 2D-periodic array of nano-rods. 
The material corresponds to silicon, with a complex refractive index 
$n=5.0 + 4.16i$ at the given wavelength obtained from tabulated data~\cite{Palik1985}.
The superspace is modeled as air, with refractive index $n=1.0$.
The parameterized geometry is schematically shown in Fig.~\ref{schematics_nanopost}, and the 
chosen geometrical parameter set is as defined in Table~\ref{table_specs_nanopost}.

\begin{table}[b]
\begin{center}
\begin{tabular}{|l|l|}
\hline
$p_x=p_y$ & 100\,nm\\ \hline
CD$_1$ / CD$_2$ / CD$_3$ & 50\,nm / 55\,nm / 60\,nm\\\hline
$h_1$ /$h_2$ & 40\,nm / 20\,nm\\ \hline
\end{tabular}
\caption{Parameter settings for the Si nano-post array (compare Fig.~\ref{schematics_nanopost}).
}
\label{table_specs_nanopost}
\end{center}
\end{table}

\begin{figure}[b]
\begin{center}
\psfrag{kx}{\sffamily \large $k_x/k_{\textrm tot}$}
\psfrag{R}{\sffamily R$_\textrm{S}$, R$_\textrm{P}$}
\psfrag{Rs (Schur)}{\sffamily R$_\textrm{S}$ (Schur)}
\psfrag{Rp (Schur)}{\sffamily R$_\textrm{P}$ (Schur)}
\psfrag{Rs (standard FEM)}{\sffamily R$_\textrm{S}$ (standard FEM)}
\psfrag{Rp (standard FEM)}{\sffamily R$_\textrm{P}$ (standard FEM)}
\includegraphics[width=.6\textwidth]{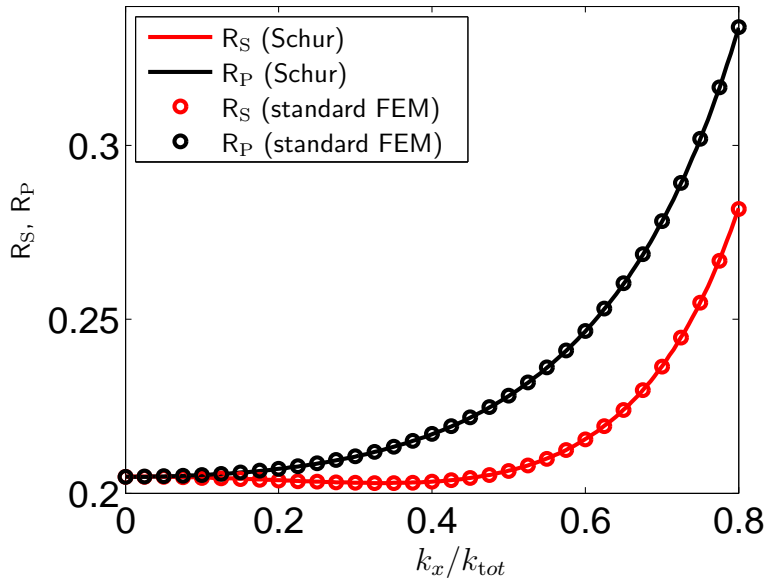}
  \caption{
Validation of the Schur complement method by comparison to standard FEM.
Angular reflection spectra for S- and P-polarized light incident on the 
nanorod array, R$_\textrm{S}$, R$_\textrm{P}$. 
The solid lines show results computed with the Schur complement method (for 322 sources). 
The open circles show results computed with standard FEM for a subset of sources. 
}
\label{fig_angle_spectrum_1d}
\end{center}
\end{figure}

First, the Schur complement method is validated: 
Angular reflection spectra of S- and P-polarized light are computed using the Schur complement method, 
using 161 directions of incidence. 
Figure~\ref{fig_angle_spectrum_1d} shows the obtained spectra. 
We validate the results by computing the same spectra using standard FEM, i.e., using separate computations for each 
angle of incidence and not decomposing the matrix $A$ into $\theta$-dependent and $\theta$-independent parts (see Eq.~\ref{eqSplit2}). 
These spectra are also plotted in Figure~\ref{fig_angle_spectrum_1d}. 
The corresponding results agree quantitatively to a level of a relative error below 0.2\%
at the chosen accuracy setting.
Here, the precision parameter for the $hp$ refinement of the finite element discretization 
was chosen as $p_{\textrm Prec}=4\times10^{-3}$, yielding a discrete problem with around $10^5$ unknowns. 
Please see a previous publication for details on the numerical settings for $hp$-refinement\cite{Burger2015al}.
Agreement of results obtained with the Schur complement method and with standard FEM is expected, however, small deviations 
of the numerical results within the ranges of numerical discretization errors 
are due to differences in numerical treatment of transparent boundaries in the Schur complement 
and in the standard FEM case.

\begin{figure}[t]
\begin{center}
\psfrag{kx}{\sffamily $k_x/k_{\textrm tot}$}
\psfrag{ky}{\sffamily $k_y/k_{\textrm tot}$}
\psfrag{RS}{\sffamily (a)}
\psfrag{RP}{\sffamily (b)}
\includegraphics[width=.48\textwidth]{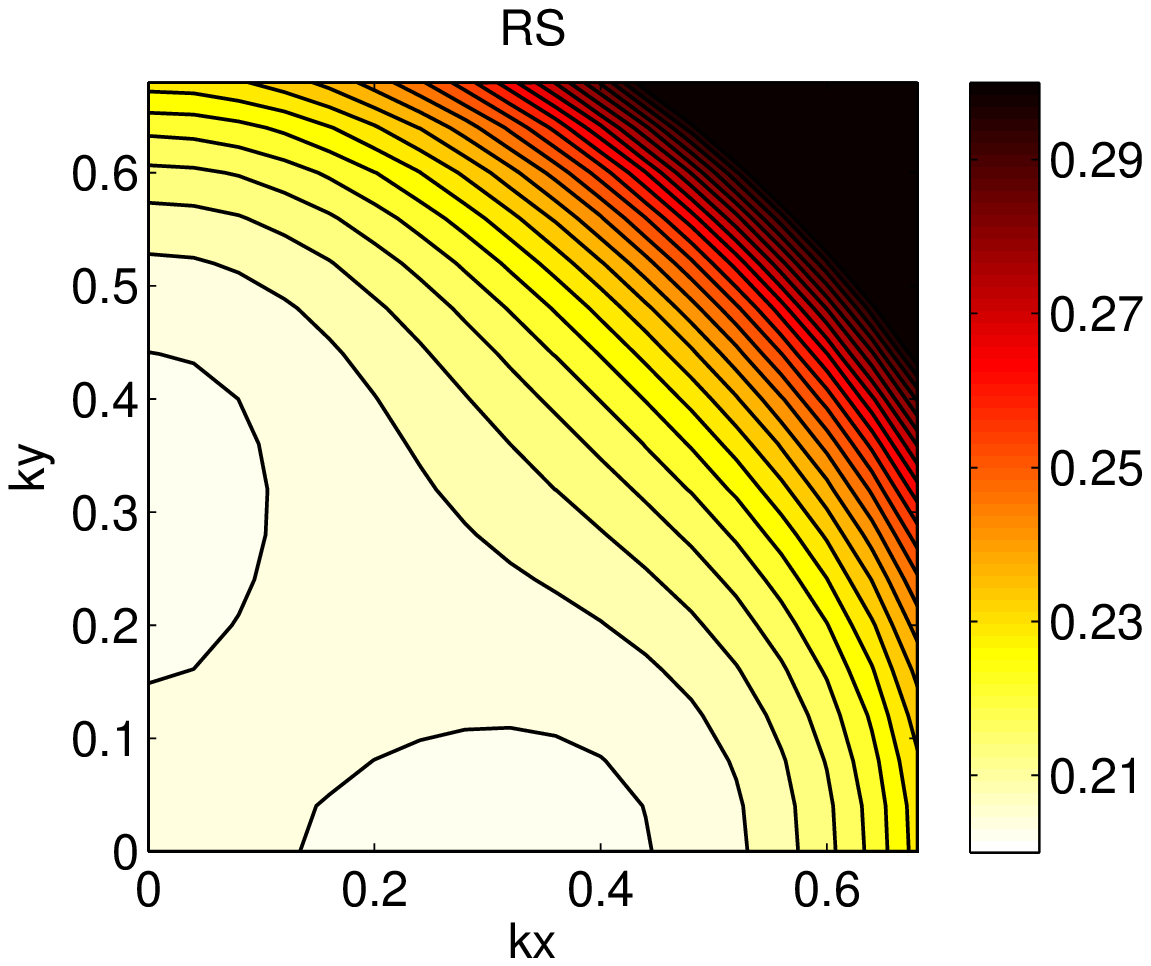}
\includegraphics[width=.48\textwidth]{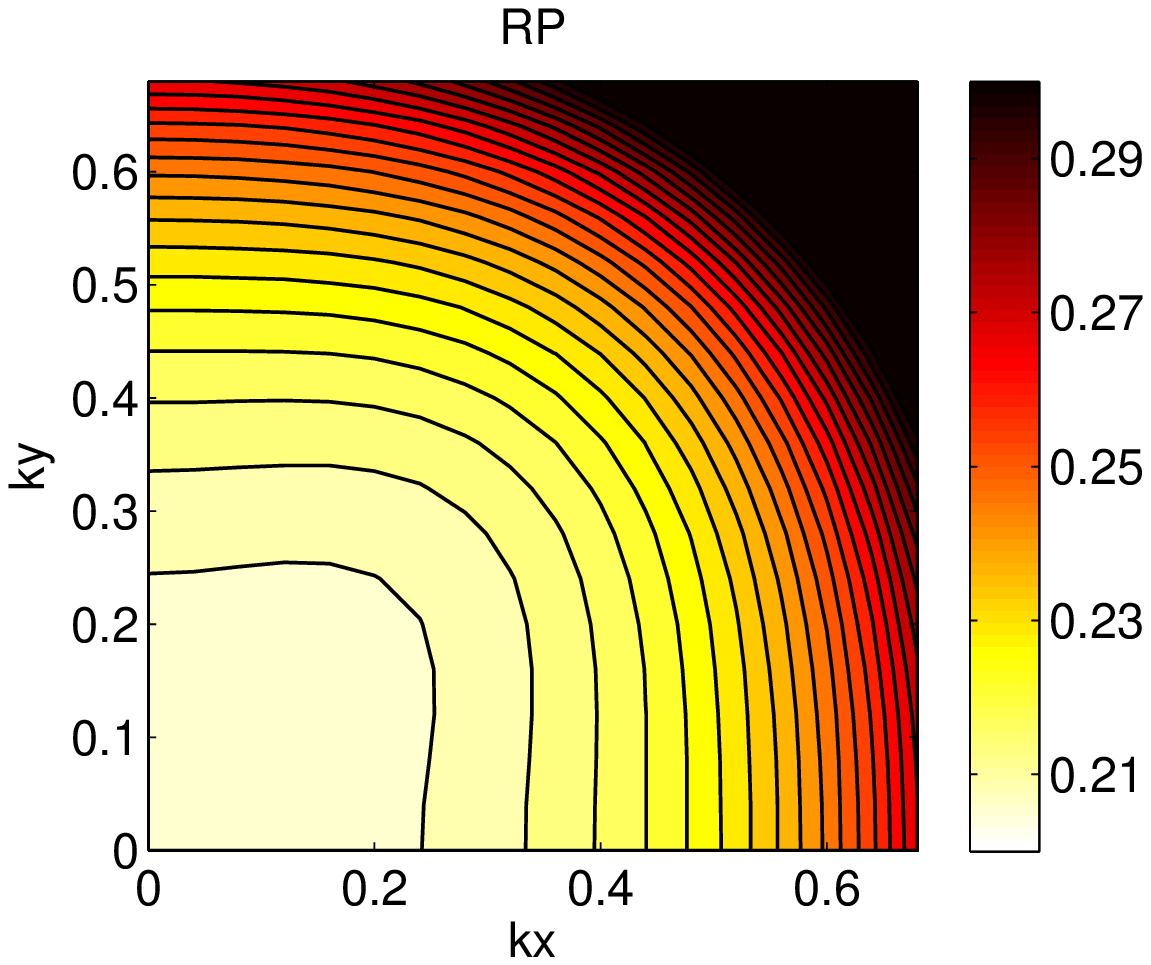}
  \caption{
Angle-resolved reflection spectra for S-polarized illumination (a) and P-polarized 
illumination (b). 
Both spectra have been obtained in a single simulation run using the Schur complement method
for a total of 848 incident source fields. 
}
\label{fig_angle_spectrum_2d}
\end{center}
\end{figure}

Next, usability of the method is demonstrated by computing two angle-resolved reflection spectra for the array, where 
both, the inclination angle and the rotation angle are varied. 
Figure~\ref{fig_angle_spectrum_2d} visualizes the resulting spectra. 
In the figure, the absolute values of the reflection coefficients for S and P polarized light are plotted 
as function of $k_x$ and $k_y$, which are the transversal components of the incident wave vector $\vec{k}=(k_x, k_y, k_z)$.
These results have been obtained in a single computation using the Schur complement method,
where  848 plane waves with different wave vectors and 
polarizations have been used. 
This results demonstrates practicability of the method for simulation projects with many independent source terms.

\begin{figure}[t]
\begin{center}
\psfrag{t [a.u.]}{\sffamily $t_{\textrm CPU}\, [a.u.]$}
\psfrag{N}{\sffamily $\textrm{N}_{\textrm src}$}
\psfrag{standard FEM}{\sffamily standard FEM}
\psfrag{Schur}{\sffamily Schur}

\includegraphics[width=.6\textwidth]{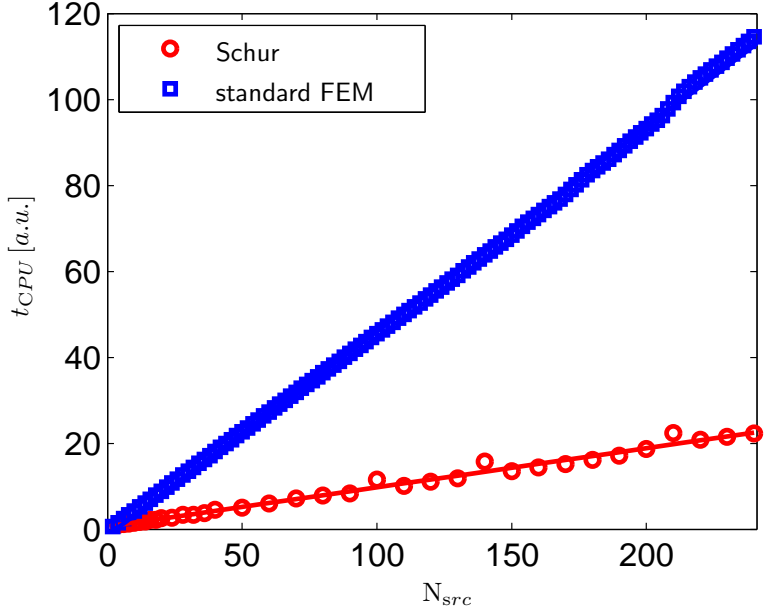}
  \caption{
Comparison of computation times for simulations of the same setup using 
the Schur complement method (red circles, linear interpolation: red line)
and using standard FEM (blue squares, linear interpolation: blue line):
Computation time $t_{\textrm CPU}$ as function of number of sources $\textrm{N}_{\textrm src}$.   
}
\label{fig_schur_cc}
\end{center}
\end{figure}

As commented above, the computational costs of a Schur method simulation run with $N$ sources 
can be significantly smaller than the total computational cost of $N$ independent simulation runs.
To demonstrate this for the given example, the CPU times for computations with both approaches are 
compared (Schur method CPU times and standard FEM CPU times are normalized with the same constant). 
Figure~\ref{fig_schur_cc} shows how the computation times scale with the number of sources. 
A speed-up by a factor of $>5$ is observed for this specific example. 
Both simulation runs have been performed with the same numerical accuracy settings (meshing parameters, 
hp-FEM discretization parameters), and on the same hardware (using a single thread of a standard computer). 
Please note that for the standard FEM computations we have also ``re-used'' the inverted system matrix 
$A^{-1}$ for the second source polarization setting (for each given wave vector). Accordingly, without 
this improved standard FEM, the speed-up factor would be $>10$ instead of $>5$.
For a comparison of standard FEM performance in comparison to other methods for solving 
Maxwell's equations we refer to previous benchmarks~\cite{Maes2013oe,Hoffmann2009spie,Burger2005bacus}.

In general, the Schur method speed-up factor obtained for a specific setup is expected to 
increase with the ratio of the sizes of the matrix blocks $A_{11}$ and $A_{22}$ 
(Eq.~\eqref{eqBlocks}). 
However, inverting the Schur complement requires inversion of a full matrix, such that 
the speed-up factor can be worsened for large problem sizes. 
Clearly, the $N$ computations for the standard (``no Schur'') simulations can also be performed 
in parallel on a multi-core processor, thus decreasing computation time by a factor of $1/N$. 
However, in typical applications where many spectra have to be computed, the 
computations using Schur complements can be parallelized in the same manner.



\section{Conclusion}
A finite-element method for computing solutions to the Maxwell 3D scattering problem has 
been presented, which makes use of a Schur complement solver for efficiently 
handling setups with many independent source terms for the same geometrical 
setup. 
A speed-up factor of greater than five has been observed for a specific 
application example in which angle-resolved reflection spectra from an 
array of silicon nano-posts are computed. 
We expect that this method can be used effectively for a variety of applications 
including computational metrology, computational lithography, nano-optics design 
in integrated optoelectronics, photovoltaics, light-emitters, and other fields. 

\section*{Acknowledgments}
We acknowledge support of BMBF through projects 13N13164 (SolarNano) and 13N12438 (MOSAIC).

\bibliography{/home/numerik/bzfburge/texte/biblios/phcbibli,/home/numerik/bzfburge/texte/biblios/my_group,/home/numerik/bzfburge/texte/biblios/lithography,/home/numerik/bzfburge/texte/biblios/jcmwave_third_party}

\bibliographystyle{spiebib}  

\end{document}